\documentclass[12 pt]{article}
\usepackage[dvips]{graphicx}
\usepackage{lscape}
\begin{document}
  \title{New Cyclic Voltammetry Method for Examining Phase
         Transitions: Simulated Results}
  \author{
    \centerline{ I. Abou Hamad$^{1,2,3}$, 
      D.T. Robb$^{3,}$\footnote{Present address: Department of Physics, Box 5721, Clarkson University, Potsdam, NY 13699, USA},
      P.A. Rikvold$^{2,3,4,}$\footnote{Corresponding author at:
      Department of Physics, Florida State University, Tallahassee, FL
      32306-4350, USA
      {\it E-mail address:\/} rikvold@scs.fsu.edu}}\\ 
    \centerline{\scriptsize\it $^{1}$HPC$^{2}$, Center for Computational Sciences, Mississippi State University, Mississippi, MS 39762-5167, USA}\\
    \centerline{\scriptsize\it $^{2}$Center for Materials Research and
    Technology and Department
    of Physics, Florida State University, Tallahassee, FL 32306-4350, USA}\\
    \centerline{\scriptsize\it $^{3}$School of Computational Science,
      Florida State University, Tallahassee, FL 32306-4120, USA}\\
    \centerline{\scriptsize\it $^{4}$National High Magnetic Field
    Laboratory, Tallahassee, FL 32310}\\
  }
  \maketitle
%  \clearpage

  \begin{abstract}
    We propose a new experimental technique for cyclic voltammetry, based on
    the first-order reversal curve (FORC) method for analysis of systems
    undergoing hysteresis. The advantages of this electrochemical FORC
    (EC-FORC) technique are
    demonstrated by applying it to dynamical models of electrochemical
    adsorption. The method can not only differentiate
    between discontinuous and continuous phase transitions, but can also
    quite accurately recover equilibrium behavior from dynamic analysis 
    of systems with a continuous phase transition. 
    Experimental data for EC-FORC analysis could easily be obtained
    by simple reprogramming of a potentiostat designed for conventional 
    cyclic-voltammetry experiments.
    \end{abstract}
  
      
      \section{Introduction}
      \label{sec:I}
      Recent technological developments in electrochemical deposition
      have made possible experimental studies of atomic-scale
      dynamics \cite{Tansel:06}. It is therefore now both timely and
      important to develop new methods for computational analysis
      of experimental adsorption dynamics. In this paper we apply one
      such analysis technique, the first-order reversal curve (FORC) method,
      to analyze model electrosorption 
      systems with continuous and discontinuous phase
      transitions. We propose that this electrochemical FORC (EC-FORC)
      method can be a useful
      new experimental tool in surface electrochemistry.
      
      The EC-FORC method was originally conceived \cite{kn:mayergoyz86} in
      connection with magnetic hysteresis. It
      has since been applied to a variety of magnetic systems, ranging
      from magnetic recording media and nanostructures to
      geomagnetic compounds, undergoing \textit{rate-independent} (i.e.,
      very slow) magnetization reversal \cite{kn:pike99}. Recently, there have
      also been several FORC studies of \textit{rate-dependent}
      reversal \cite{kn:enachescu05,kn:fecioru-morariu04,kn:robb05}.
      Here we introduce and apply the FORC method in an electrochemical
      context. 
      
      The FORC analysis is applied
      to rate-dependent adsorption in two-dimensional
      lattice-gas models of electrochemical deposition. We study the
      dynamics of two specific models, using kinetic Monte Carlo (KMC)
      simulations. First, we consider 
      a lattice-gas model with attractive nearest-neighbor
      interactions (a simple model of underpotential deposition, UPD),
      being driven across its discontinuous phase transition by a
      time-varying electrochemical potential. Second, we study
      a lattice-gas model with repulsive lateral interactions and
      nearest-neighbor exclusion (similar to the model of halide adsorption on
      Ag(100), described in
      Refs.~\cite{AbouHamad:04,MitchellSS:01,AbouHamad:03,AbouHamad:05}), being
      similarly driven through its continuous phase transition.
      Some preliminary results of this work have been submitted for
      publication elsewhere \cite{AbouHamad:06}.
      
      The rest of this paper is organized as follows. In Sec.~\ref{sec:forc}
      the FORC method is explained. The lattice-gas model used both for
      systems with continuous and discontinuous transitions is
      introduced in Sec.~\ref{sec:M}. In Sec.~\ref{sec:S} the dynamics of
      systems with a discontinuous phase transition are studied. The
      dynamics of systems with a continuous phase transition are studied
      in Sec.~\ref{sec:C}. Finally, a comparison between
      the two kinds of phase transitions is presented
      together with our conclusions in Sec.~\ref{sec:conc}.
      
      \section{The EC-FORC Method} 
      \label{sec:forc}
\begin{figure}
	\begin{center}
	  \includegraphics[width=.4\textwidth]{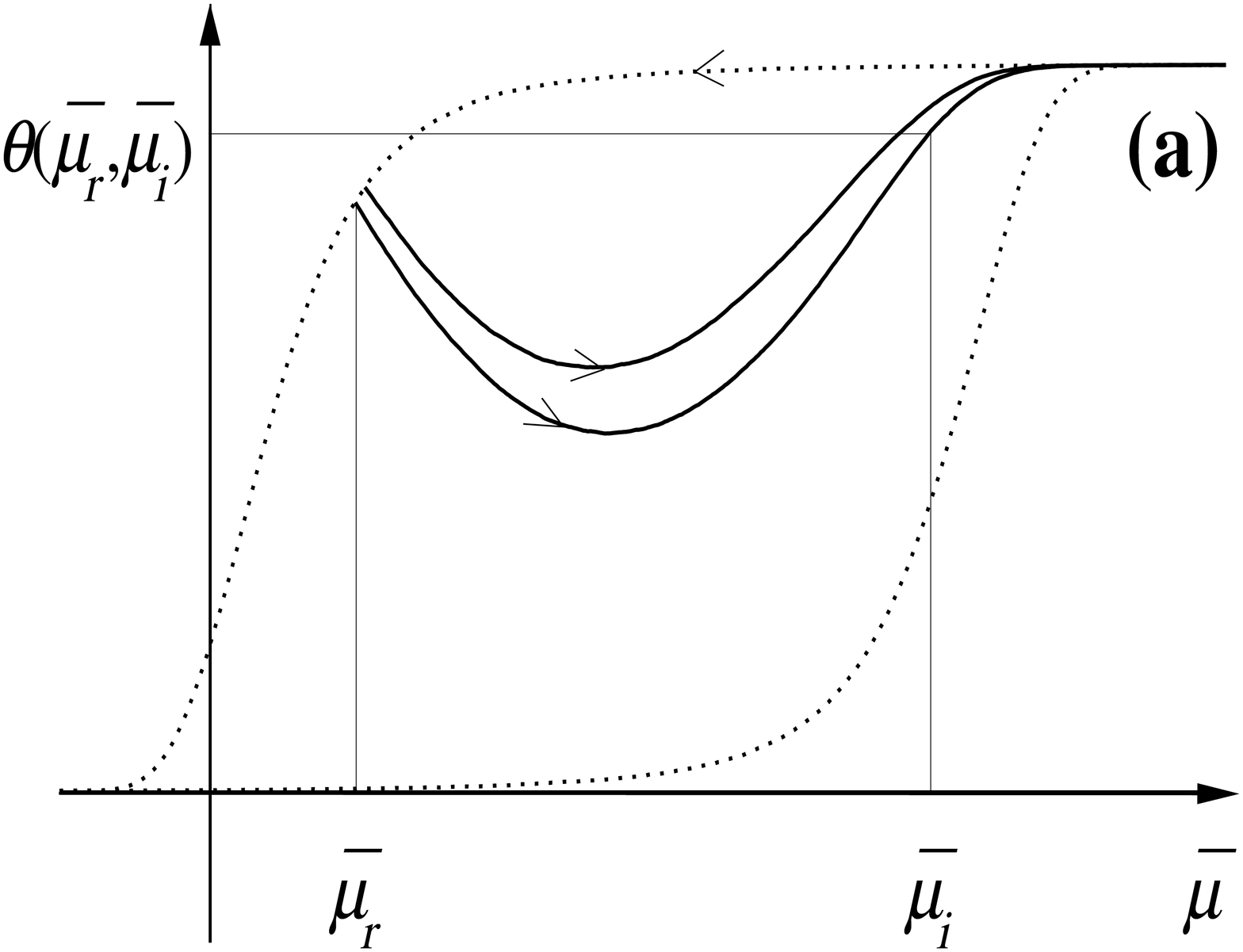}
	  \hspace{0.5truecm}
	  \includegraphics[width=.4\textwidth]{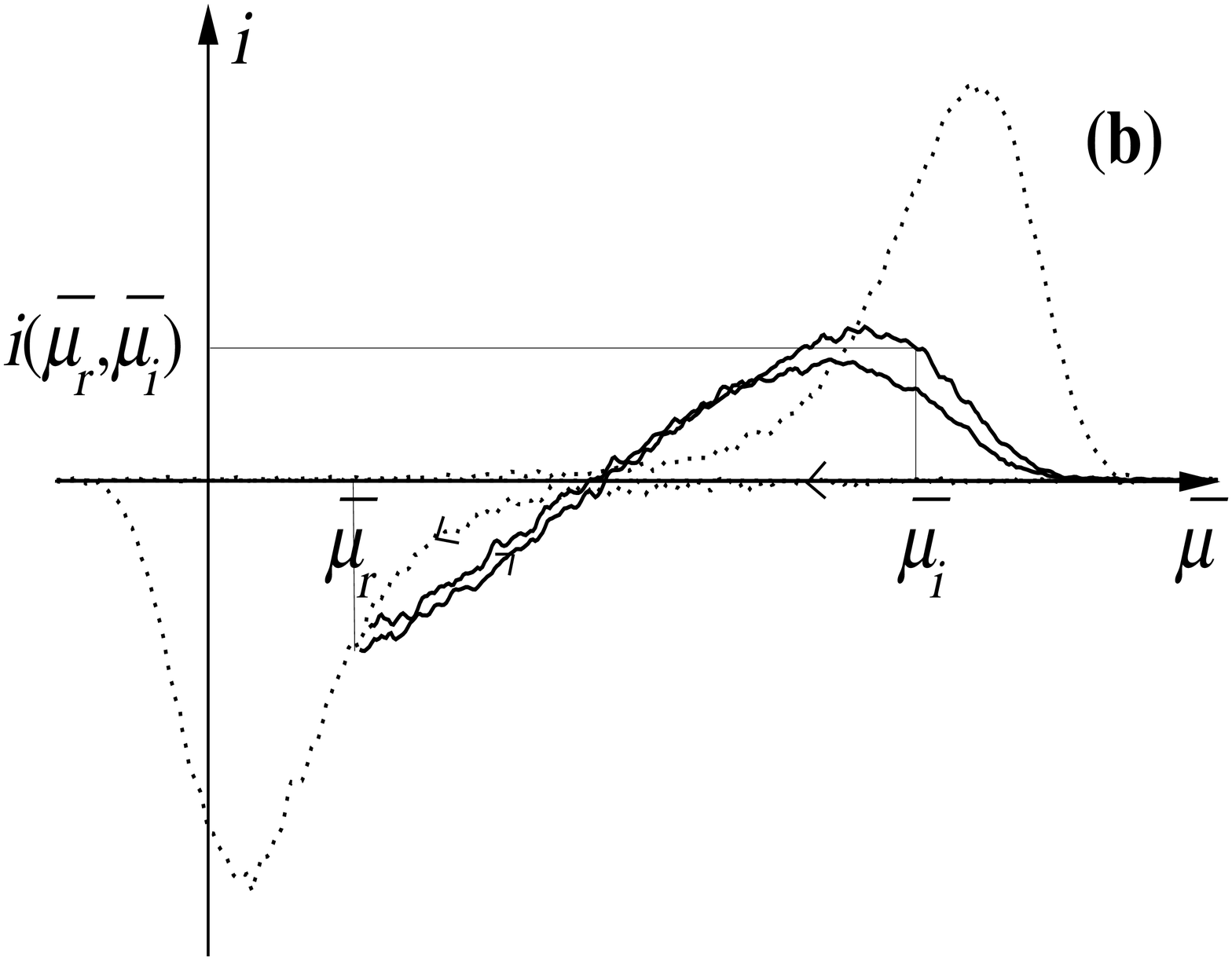}
	\end{center}
	  \caption[]{({\bf a}) Schematic diagram of
            two first-order reversal curves (FORCs) 
	    $\theta(\bar{\mu}_r, \bar{\mu}_i)$, separated by 
	    the reversal-field step size $\Delta \bar{\mu}_r$ (solid lines). 
	    The dotted lines represent the major hysteresis loop.
	    ({\bf b}) The corresponding voltammetric currents, 
	    $i(\bar{\mu}_r, \bar{\mu}_i) \propto \partial \theta /
	    \partial \bar{\mu}_i$.}
	  \label{fig:loop}
      \end{figure}

      For an electrochemical adsorption system, the FORC method consists of
      saturating the adsorbate coverage $\theta$ in a strong positive
      electrochemical potential $\bar{\mu}$ 
      (proportional to the electrode potential, with the same sign for
      anions and opposite sign for cations) and, in each case starting from
      saturation, decreasing $\bar{\mu}$ at a constant scan rate $\Omega$
      to a series of progressively
      more negative ``reversal potentials'' $\bar{\mu}_r$.  
      (See Fig.~\ref{fig:loop}).
      Subsequently, the potential is increased back to the saturating
      value at the same rate $\Omega$ \cite{kn:pike99}. The method is 
      thus a simple generalization of the
      standard cyclic voltammetry (CV) method, in which the negative return
      potential is decreased for each cycle. This produces a family of
      FORCs, $\theta (\bar{\mu}_r, \bar{\mu}_i)$, where $\theta$ is the
      adsorbate coverage and $\bar{\mu}_i$ is the instantaneous
      potential during the increase back toward saturation.
Alternatively, as it is usually done in CV experiments, one can record
the family of voltammetric currents, 
\begin{equation}
i(\bar{\mu}_r,\bar{\mu}_i) 
=
- \gamma e \frac{{\rm d} \bar{\mu}_i}{{\rm d} t}
           \frac{\partial \theta }{\partial \bar{\mu}_i}
	   \;,
\label{eq:curr}
\end{equation}
where $\gamma$ is the electrosorption valency 
\cite{VETT72A,VETT72B,RIKV06r} and $e$ is the elementary charge.  
      Although we shall not discuss this further here, it is of course also
      possible to fix the negative limiting potential and change
      the positive return potential from cycle to cycle.
      
\begin{figure}
\begin{center}
\includegraphics[width=.6\textwidth]{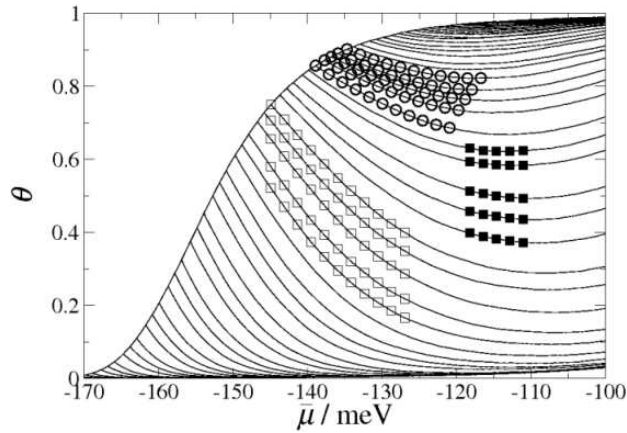}
\caption[]{A subset of the family of FORCs. 
The circled points are an example of the `slanted' $5\times11$ array of data
points used to calculate $\rho$ from $\theta(\bar{\mu}_r,\bar{\mu}_i)$. 
The filled squares represent a `square'
$5\times5$ array and the hollow squares represent a `square' $5\times11$ array. See discussion in the text.}
\label{fig:grid}
\end{center}
\end{figure}
      The family of FORCs consists of $N$ FORCs measured at 
      values of $\bar{\mu}_r$ that are evenly spaced by $\Delta
      \bar{\mu}_r$. Each of these FORCs has data points measured 
      at the (in general different)
      constant spacing $\Delta\bar{\mu}_i$ \cite{kn:pike99}. 
      It is further useful to calculate the FORC distribution,
\begin{equation}
\rho=- \frac{1}{2} \frac{\partial^2 \theta}{\partial\bar{\mu}_r\,
\partial\bar{\mu}_i}
=
\frac{1}{2 \gamma e ( {\rm d} \bar{\mu}_i / {\rm d} t )} 
\frac{\partial i(\bar{\mu}_r,\bar{\mu}_i)}{\partial \bar{\mu}_r}
	  \;, 
	\label{forc.definition}
      \end{equation}
      which measures the sensitivity of the dynamics to the progress
	of reversal along the outermost hysteresis loop or major 
	loop.\footnote{Note that to
	normalize the FORC distribution, the term $\frac{1}{2} \delta
	(\bar{\mu}_i~-~\bar{\mu}_r ) \frac{\partial \theta
	(\bar{\mu}_r,\bar{\mu}_i)}{\partial \bar{\mu}_i} |_{\bar{\mu}_i \to
	  {\bar{\mu}_r}^+}$ must be added to
	Eq.~(\ref{forc.definition}) \cite{kn:pike03}. Here we consider
	the distribution only away from the line $\bar{\mu}_i =
	\bar{\mu}_r$. The 
	additional term could be found from the major loop.}

The details of the 
calculation of $\rho$ depend on whether the measured quantity is 
the coverage $\theta(\bar{\mu}_r,\bar{\mu}_i)$, which is the most
convenient in simulation studies, or the voltammetric current 
$i(\bar{\mu}_r,\bar{\mu}_i)$, which would be most usual in CV
experiments. In the latter case, the calculation involves simple 
one-dimensional numerical differentiation with respect to $\bar{\mu}_r$, 
e.g. $\rho=[2\gamma e({\rm d}\bar{\mu}_i / {\rm d}t)]^{-1}
[i(\bar{\mu}_r+ \Delta \bar{\mu}_r,\bar{\mu}_i) -
i(\bar{\mu}_r,\bar{\mu}_i) ] / \Delta \bar{\mu}_r$.
Geometrically it is proportional to the vertical distance between the
current traces. (See Fig.~\ref{fig:loop}(b).) 
	
In the case that $\theta (\bar{\mu}_r, \bar{\mu}_i)$ is the measured
quantity, as in the simulations presented here, the calculation of
$\rho$ is somewhat more complicated. 
        An array of data points $\theta (\bar{\mu}_r, \bar{\mu}_i)$
        from consecutive FORCs is used to calculate $\rho$
	as shown in Fig.~\ref{fig:grid}. 
	A polynomial surface is fit to this array
        of data points to provide a functional form of 
	$\theta (\bar{\mu}_r, \bar{\mu}_i)$,
        the mixed partial derivative of which is proportional to $\rho$
	as given by Eq.~(\ref{forc.definition}).  
Away from the major hysteresis loop, the array of data points used to
calculate $\rho$ can be taken to be `square', i.e.,
to start and end at the same value of $\bar{\mu}_i$ for each
of the FORCs involved (squares in Fig.~\ref{fig:grid}). Close to 
the major loop, this is not
geometrically possible. As shown in Fig.~\ref{fig:grid}, we therefore used a
`slanted' array of data points to calculate $\rho$.
To obtain the values of $\rho$ at grid points other than the center point of 
the grid ($\bar{\mu}_i$, closer to $\bar{\mu}_r$) we used a  third-order 
polynomial to fit $\theta(\bar{\mu}_r, \bar{\mu}_i)$. This allows the evaluation
of the mixed partial derivative $\rho$ at all points of the grid, while a second-order polynomial fit only allows the evaluation of $\rho$ at the center point. 
For consistency and convenience, a third-order polynomial was used with the 
three kinds of grids described above and at all points close to and away from 
the major loop.
The three kinds of
grids were compared for points away from the major hysteresis loop. While
the larger $5\times11$ grids (`square', and `slanted') produced very similar 
and smooth FORC diagrams,
the FORC diagram produced by the $5\times5$ grid was more noisy. Consequently,
in the results shown in this paper, the $5\times11$ `slanted' grid was used to 
calculate $\rho$ from $\theta(\bar{\mu}_r, \bar{\mu}_i)$ 
for all values of the arguments. 

Another approach to calculating $\rho$ near the major loop would be to use an
`extended FORC diagram' \cite{kn:pike03}, for
which a square grid of points can be used even near the major loop,
but where the delta function normalization term
(See Footnote 1) will automatically be included by the polynomial fitting procedure described above. The methods for calculating $\rho$ that are described
here, are quite straightforward, but also rather slow. A faster method
%, which may well be useful in electrochemical applications, 
is based on the Savitzky-Golay smoothing algorithm~\cite{Savitzky:64,Press:nr}. It is described in Ref.~\cite{Helsop:05}.

The FORC distribution is usually displayed as a contour plot called a
`FORC diagram.' A positive value of $\rho$ indicates that the
corresponding FORCs are converging with increasing
$\bar{\mu}_i$, while a negative value indicates divergence. 
The physical significance of the sign of $\rho$ will become clearer
when we apply the EC-FORC method to discontinuous and continuous phase
transitions in Secs.~\ref{sec:S} and ~\ref{sec:C}, respectively.

	\section{Model}
	\label{sec:M}
	KMC simulations of lattice-gas models, where a
	Monte Carlo step (MCS) corresponds to one attempt to cross a
	free-energy barrier, have been used to simulate the kinetics of
	electrochemical adsorption in systems with both
        discontinuous \cite{AbouHamad:03,AbouHamad:05,Frank:05,Frank:06}
	and continuous \cite{AbouHamad:04,Mitchell:02} phase transitions. 
	The energy associated with a lattice-gas configuration is
	described by the grand-canonical effective Hamiltonian for an
	$L\times L$ square system of adsorption sites,
	\begin{equation}
	  {\mathcal{H}} = - \sum_{i<j} \phi_{ij} c_{i} c_{j} - \bar{\mu} 
	  \sum_{i=1}^{L^2}c_{i}\; ,
	  \label{eq:H-II}
	\end{equation}
	where $\sum_{i<j}$ is a sum over all pairs of sites, $ \phi_{ij} $ are
	the lateral interaction energies between particles on the $i$th and
	$j$th sites measured in meV/pair, and $ \bar{ \mu } $ is the
	electrochemical potential measured in meV/atom. The local
	occupation variables $ c_{i} $ can take the values 1 or 0, depending
	on whether site $ i $ is occupied by an ion (1) or solvated (0).
	The sign convention is chosen such that $\bar\mu > 0$ favors
	adsorption. Negative values of $\phi_{ij}$ denote repulsion,
	while positive values of $\phi_{ij}$ denote 
	attraction between adsorbate particles on the surface. In addition
	to adsorption/desorption steps, we include diffusion steps
	that have a free-energy barrier comparable to the adsorption/desorption
        free-energy barrier \cite{AbouHamad:04}. 
	
	The dynamics of the model are studied by a KMC simulation with
	the computational time unit of one Monte Carlo Step per Site (1
	MCSS = $L^2$ MCS). 
	For a discussion of the relation between this simulated
	time unit and real, physical time, see Ref.~\cite{AbouHamad:04}. 
	In each MCS of the simulation, an adsorption site is
	chosen at random and a move (adsorption, desorption, or diffusion)
        is attempted.
	The transition rates from the present configuration to the set of new
	possible configurations are calculated. A weighted
	list of the probabilities for accepting each of these moves 
	during one MCS is constructed using
	Eq.~(\ref{eq:P}) below, and used to calculate the probabilities
	$R\rm{(F|I)}$ of the individual moves between the initial state $\rm I$
	and final state $\rm F$. The probability for
	the system to remain in the initial configuration at the end of
	the time step is consequently $R\rm{(I|I)}=1-$$\Sigma_{\rm F\neq
	I}R\rm{(F|I)}$ \cite{AbouHamad:04,MitchellSS:01}.
	
	Using a thermally activated, stochastic
	barrier-hopping picture, the energy of the transition state for a
	microscopic change from an initial state $\rm I$ to a final state
	$\rm F$ is approximated by the symmetric Butler-Volmer
	formula \cite{Brown:99,Kang:89,Buendia:04}
	\begin{equation}
	  U_{\rm T_{\lambda}}=\frac{U_{\rm I}+U_{\rm
	F}}{2}+\Delta_{\lambda} \; .
	\end{equation}
	Here $U_{\rm I}$ and $U_{\rm F}$ are the energies of the
	initial and final states as given by Eq.~(\ref{eq:H-II}), 
	$\rm T_{\lambda}$ is the transition state for process $\lambda$, 
	and $\Delta_{\lambda}$ is a ``bare'' barrier associated with
	process $\lambda$. This process can here be either
	nearest-neighbor diffusion ($\Delta_{\rm nn}$),
	next-nearest-neighbor diffusion ($\Delta_{\rm nnn}$), or
	adsorption/desorption ($\Delta_{\rm a/d}$). The probability per
	unit time for a particle to make a transition from state $\rm I$ to
	state $\rm F$ is approximated by the one-step Arrhenius
	rate \cite{Brown:99,Kang:89,Buendia:04}
	\begin{equation}
	  \mathcal{R}({\rm F}|{\rm I})= 
	  \nu \exp\left(-\frac{U_{\rm T_{\lambda}}-U_{\rm I}}
	  {k_{\rm B}T}\right) =  \nu \exp
	  \left(-\frac{\Delta_{\lambda}}{k_{\rm B}T}\right)
	  \exp\left(-\frac{U_{\rm F}-U_{\rm I}}{2k_{\rm
	  B}T}\right),\label{eq:P}
	\end{equation}
	where $k_{\rm B}$ is Boltzmann's constant and $T$ is the
	absolute temperature. 
	Here, $\nu$ is the attempt frequency, which sets the overall
	timescale for the simulation. We set $\nu$ equal to 1
	MCSS$^{-1}$, so that the transition probabilities in a single
	time step (MCS) are given by 
	$R({\rm F}|{\rm I}) = \mathcal{R}({\rm F}|{\rm I}) L^{-2}$~MCSS. 
	The electrochemical potential $\bar \mu$ is changed  
	each MCSS, preventing the system from reaching equilibrium at the
	instantaneous value of $\bar \mu$.
	
	Independent of the diffusional degree of freedom, attractive
	interactions ($\phi_{ij}>0$) produce a discontinuous phase transition
	between a low-coverage phase at low $\bar\mu$, and a high-coverage
	phase at high $\bar\mu$. In contrast, repulsive interactions
	($\phi_{ij}<0$) produce a continuous 
	phase transition between a low-coverage disordered phase
	for low $\bar\mu$, and a high-coverage, ordered phase for high
	$\bar\mu$. Examples of systems with a discontinuous phase
	transition include underpotential
	deposition \cite{Frank:05,Frank:06,BrownJES:99}, while the
	adsorption of halides on Ag(100) 
	\cite{AbouHamad:04,MitchellSS:01,Mitchell:02,Ocko:97,Wandlowski:01}
	are examples of systems with a continuous phase transition.

	\section{Discontinuous Phase Transition}
	\label{sec:S}
\begin{figure}
	  \begin{center}
           \includegraphics[width=.45\textwidth]{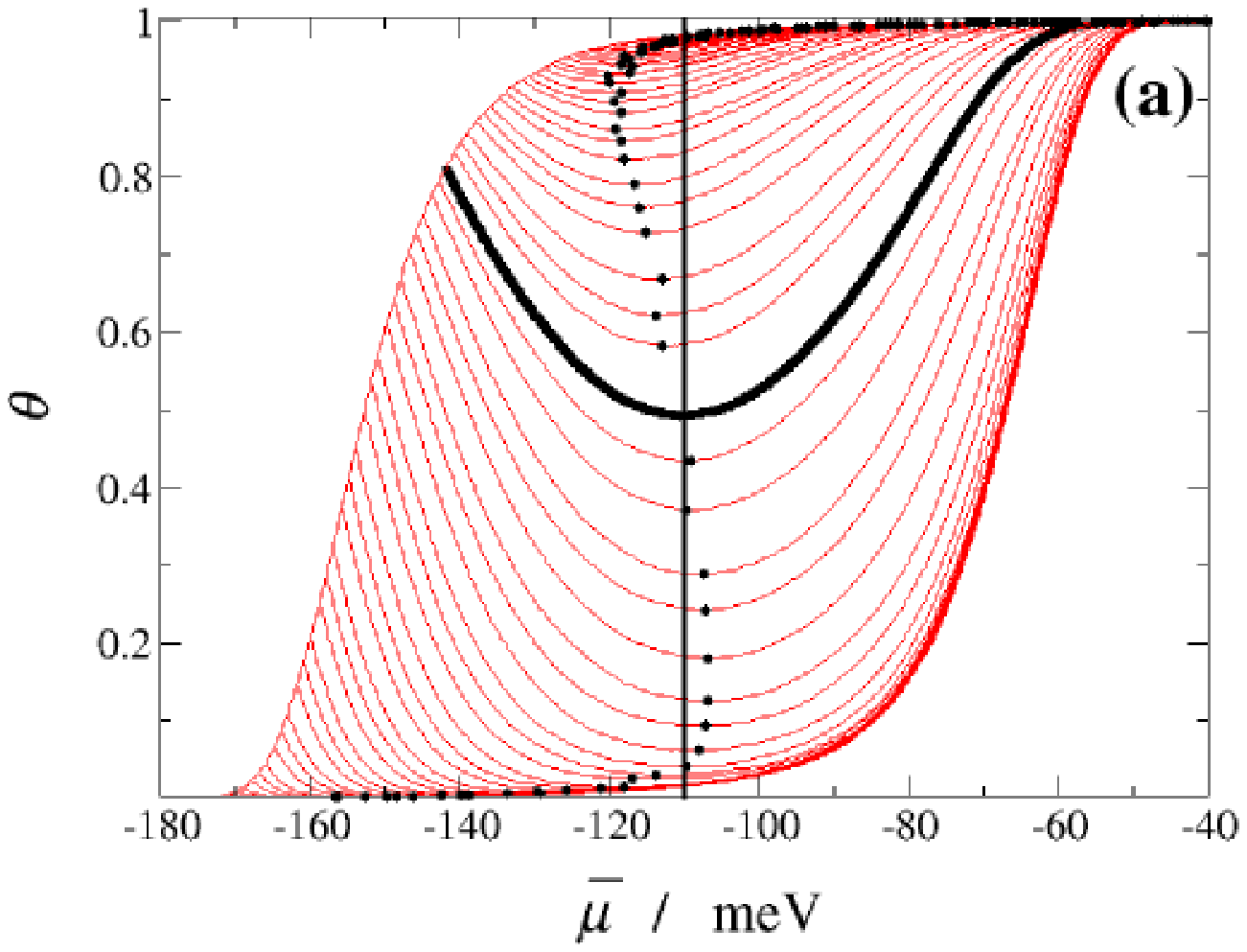} 
	   \hspace{0.1truecm}
	   \includegraphics[width=.50\textwidth,angle=0]{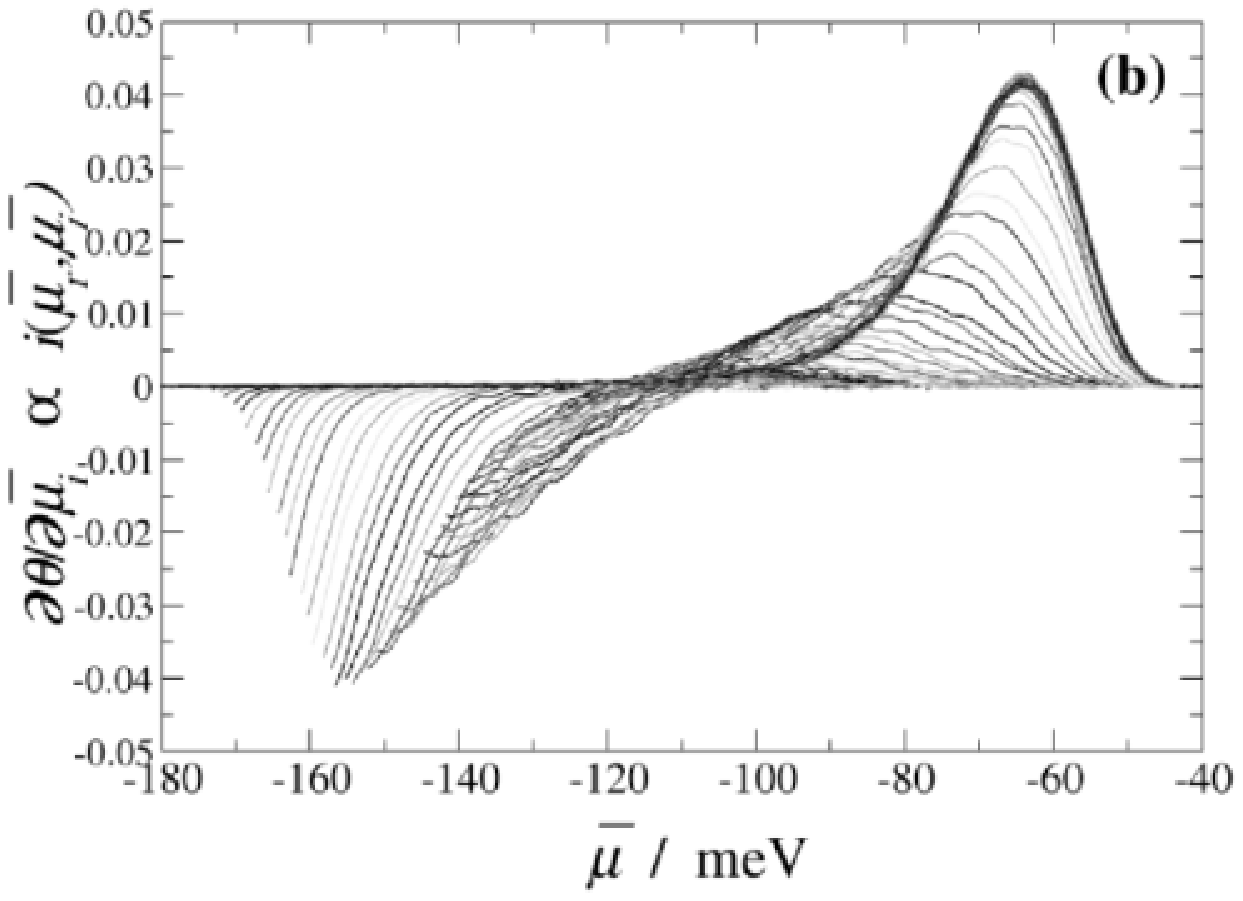}
	   \mbox{}
	   \vspace{0.5truecm}
	   \mbox{}
	    \includegraphics[width=.5\textwidth]{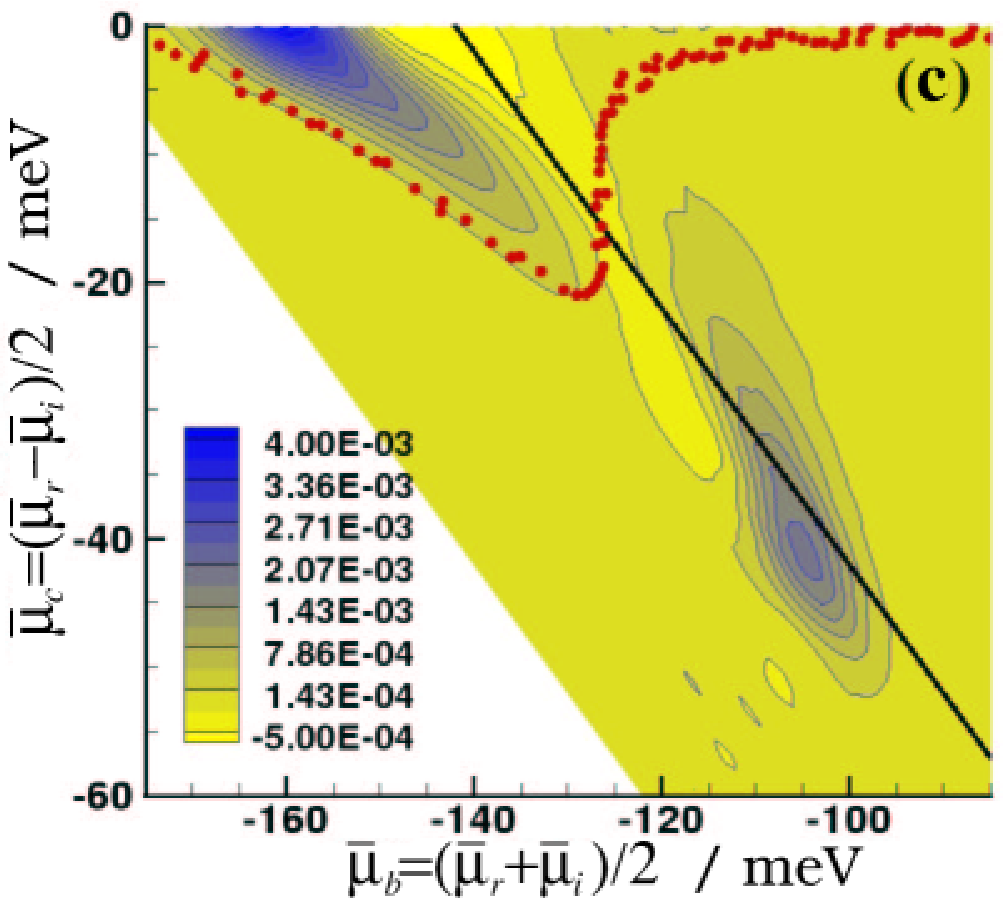}
	  \end{center}
	  \caption[]{ (Color online.)
	    ({\bf a}) FORCs for a discontinuous
	    phase transition, obtained with scan rate $\Omega =
	    0.03$~meV/MCSS and interactions and barrier heights as given
	    in the text. The vertical line shows 
	    the coexistence value of the electrochemical potential, 
	    $\bar{\mu} = \bar{\mu}_0$. The
	    minimum  of each FORC is also shown (filled circles). 
	    The thick curve corresponds to the FORC whose minimum 
	    lies nearest the coexistence value, $\bar{\mu}_i = \bar{\mu}_0$. 
({\bf b})
The corresponding voltammetric currents, calculated by
numerical differentiation of the FORCs. See Eq.~\ref{eq:curr}. 
	    ({\bf c})
	    FORC diagram generated from the family of FORCs in
	    ({\bf a}). The positions of the FORC minima are shown
	    as filled circles. The thick, 
	    straight line corresponds to the FORC marked as a 
	    thick curve in ({a}).  
%After Ref.~\protect\cite{AbouHamad:06}.
	  }
	  \label{fig:fig1}
	\end{figure}

	A two-dimensional lattice gas with attractive adsorbate-adsorbate
	lateral interactions that cause a discontinuous phase transition
	is a simple model of electrochemical underpotential
	deposition \cite{Frank:05,Frank:06,BrownJES:99,Ramos:99}. 
	Using a lattice-gas model with attractive interactions on an
        $L \times L$ lattice with $L=128$, a
	family of FORCs were simulated, averaging over ten
	realizations for each FORC at room temperature.
	The lateral interaction energy (restricted to nearest-neighbor)
	was taken to be $\phi_{ij}= \phi_{\rm nn} = 55 $\,meV, where
	the positive value indicates nearest-neighbor attraction. For
	this value of $\phi_{\rm nn}$, room temperature
	corresponds to $T=0.8 T_c$, where $T_c$ is the critical
	temperature. The barriers for adsorption/desorption and
	diffusion (nearest-neighbor only) were $\Delta_{\rm
	a/d}=\Delta_{\rm nn} = 150$\,meV, corresponding to relatively slow
        diffusion \cite{Frank:06}. Simulation runs with faster
	diffusion ($\Delta_{\rm nn} = 125$\,meV) and the same
	adsorption/desorption barrier were little different
	from the results shown in Fig.~\ref{fig:fig1}.
	The reversal electrochemical potentials $\bar{\mu}_r$
	associated with the FORCs were separated by 
	$\Delta \bar{\mu}_r = 1$\,meV
	increments in the interval $[-200\,{\rm meV},0\,{\rm meV}]$, and the
	field-sweep rate was constant at 
	$\Omega = |{\Delta}\bar{\mu}_i/{{\Delta}t}| = 0.03$~meV/MCSS. 
	The FORCs are shown in Fig.~\ref{fig:fig1}({a}), with a vertical
	line indicating the position of the coexistence value of the
	electrochemical potential, $\bar{\mu}_0=-110$\,meV, and a filled circle
	showing the position of the minimum of each FORC.
The corresponding voltammetric currents are shown in
Fig.~\ref{fig:fig1}({b}). 
	
	In a simple Avrami's-law analysis, the FORC minima all lie at
	$\bar{\mu}_i=\bar{\mu}_0$ \cite{kn:robb05}. However, in the
	simulations the minima are displaced. For $\theta>0.5$, the minima
	occur at $\bar{\mu}_i<\bar{\mu}_0$, precisely at the
	points where the tendency to phase-order, which drives local regions
	of the system toward the nearby metastable state ($\theta\approx1$),
	is momentarily balanced by the electrochemical potential, which
	drives the system toward the distant stable state
	($\theta\approx0$). For $\theta<0.5$, the stable and
	metastable states are $\theta\approx1$ and $\theta\approx0$,
	respectively, and the same balancing effect explains the FORC minima
	occurring at $\bar{\mu}_i>\bar{\mu}_0$.  The net effect
	is a `back-bending' of the curve of minima, as seen in
	Fig.~\ref{fig:fig1}({a}).
	\begin{figure}
	  \begin{center}
	    \includegraphics[width=.65\textwidth]{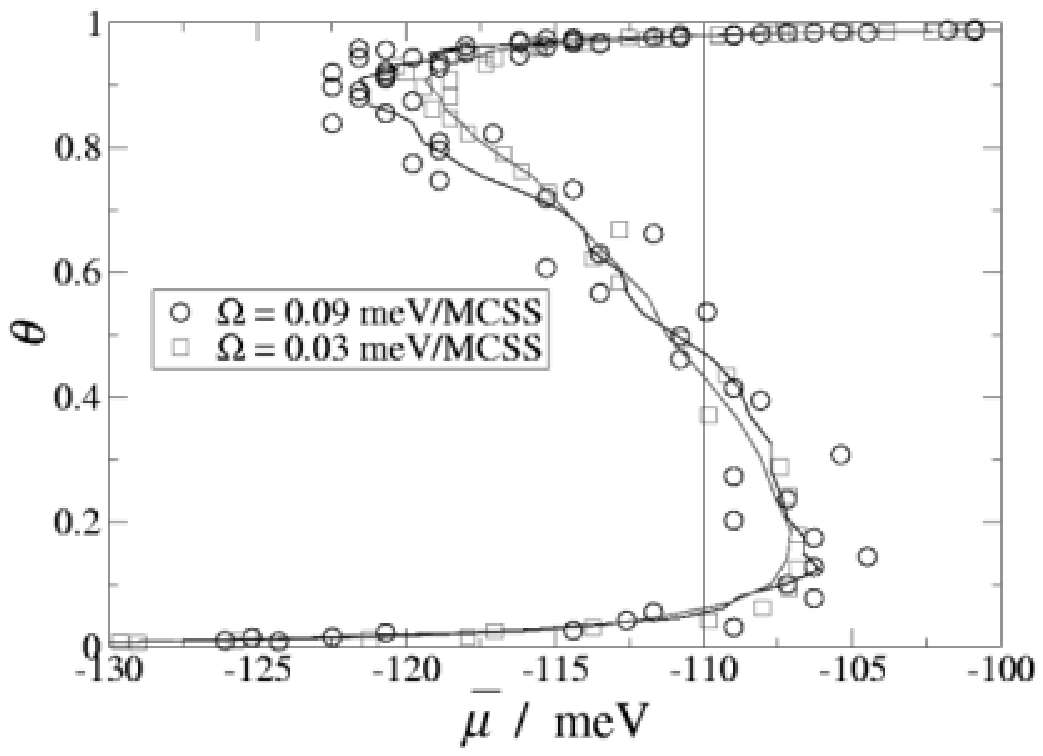}
	    \caption[]{The dependence of the FORC minima on the 
	    sweep rate $\Omega$. The figure shows FORC
	      minima for two families of FORCs with different sweep rates
	      ($\Omega=0.03$ and $0.09$ meV/MCSS, respectively). The
	      curves are guides to the eye, obtained by smoothing the
	      data using a first-order Savitzky-Golay
	      filter with a window of 5 points
	      \protect\cite{Savitzky:64,Press:nr}.
	    }
	    \label{fig:vanloop}
	  \end{center}
	\end{figure}
	
In Fig~\ref{fig:fig1}({c}), the FORC
diagram is plotted against the variables $\bar{\mu}_b =
(\bar{\mu}_r + \bar{\mu}_i)/2$  and $\bar{\mu}_c = (\bar{\mu}_r
- \bar{\mu}_i)/2$. These variables are commonly used in the
literature for plotting FORC diagrams \cite{kn:pike99}, as 
$\bar{\mu}_b$ denotes the midpoint between $\bar{\mu}_r$ and 
$\bar{\mu}_i$, and $\bar{\mu}_c$ is proportional to the
distance between these two values of $\bar{\mu}$.\footnote{In magnetic 
applications, the variables $H_b$ and $H_c$ (corresponding to $\bar{\mu}_b$ 
and $\bar{\mu}_c$, respectively) have a clear physical meaning as the
bias and coercive fields,
respectively, in a bistable magnetic system. While this physical meaning
does not extend clearly to electrochemical systems, this choice of
variables
does produce FORC diagrams with less unused space than the variables
$(\bar{\mu}_r$, $\bar{\mu}_i)$, for which the region
$\bar{\mu}_i < \bar{\mu}_r$ is forbidden,
and so we have used it here. However, the analysis described in the
following sections could also be made with FORC diagrams plotted
using the variables $(\bar{\mu}_r$, $\bar{\mu}_i)$.
}

	The definition in Eq.~(\ref{forc.definition}) implies that the FORC
	distribution $\rho$ should be negative
	in the vicinity of the back-bending. This can
	be readily seen in Fig~\ref{fig:fig1}({c}).  
	The negative values of $\rho$ reflect a local
	divergence of the FORCs away from each other as $\bar{\mu}_i$ 
	(and time) increases. This can be
	considered a dynamical instability, caused by the competition
	between the tendency to phase-order and the effect of the
	electrochemical potential. When
	the potential sweep is stopped suddenly at a potential in this
	unstable region, the subsequent time evolution of 
	$\theta$ is non-monotonic: it first approaches its metastable
	value, but then reverses and  
	relaxes reliably to its equilibrium value at that
	potential. The only exception is the
	point ($ \bar{\mu} = \bar{\mu}_0 , \theta = 0.5$) along the FORC
	indicated by bold lines in Fig.~\ref{fig:fig1}. 
	It is also interesting to note that
	the curve connecting the minima of the FORCs resembles the van
	der Waals loop in the mean-field isotherm of a fluid
	system below its critical temperature \cite{Castellan}, 
	but with an asymmetrical shape about
	the point ($\bar{\mu} = \bar{\mu}_0, \theta = 0.5$) and with
	a sweep-rate dependent shape as shown in Fig.~\ref{fig:vanloop}.
	
	\section{Continuous Phase Transition}
	\label{sec:C}
	\begin{figure}
	  \vspace{0.4truecm}
	  \begin{center}
	    \includegraphics[width=.45\textwidth]{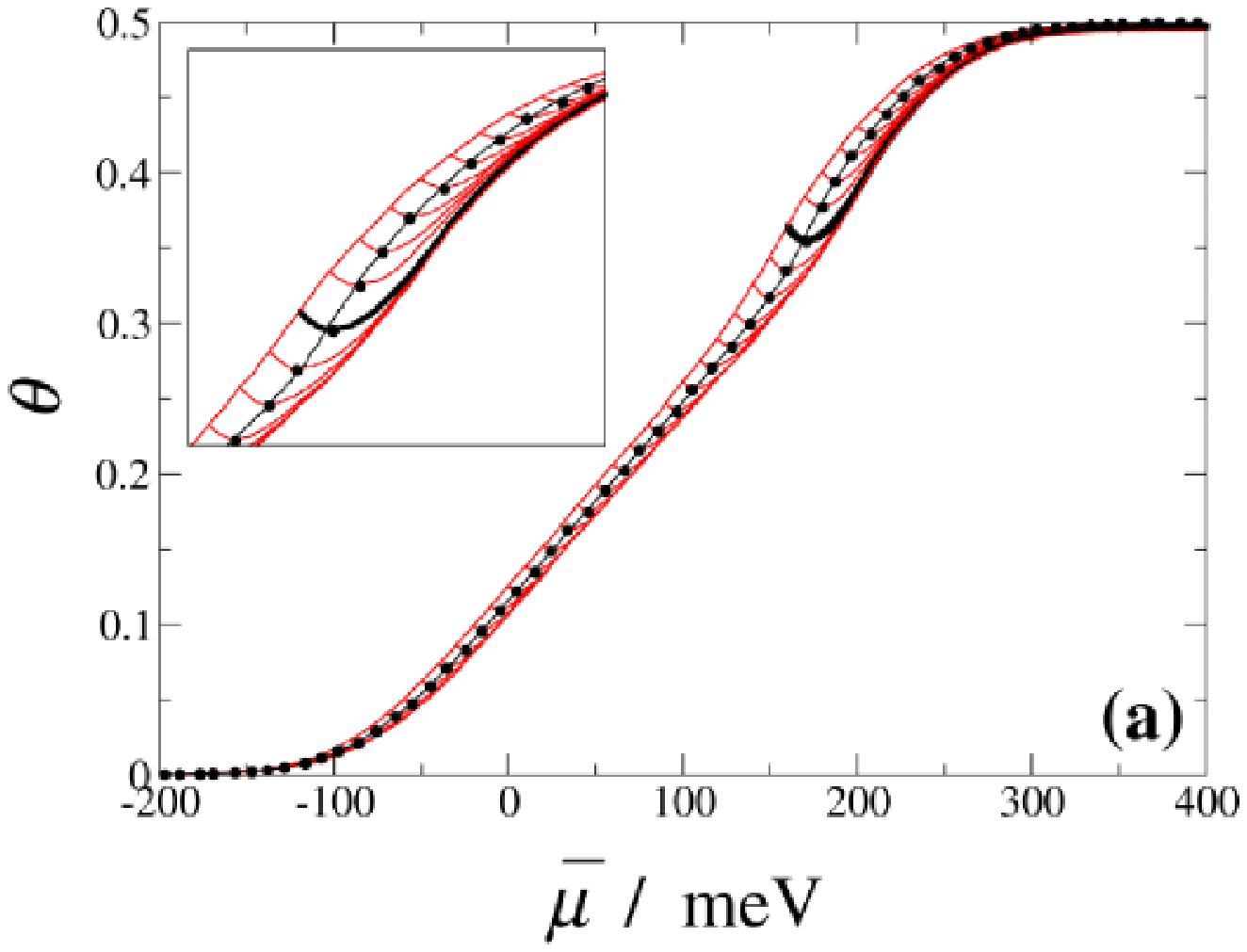}
	    \hspace{0.1truecm}
	    \includegraphics[width=.50\textwidth]{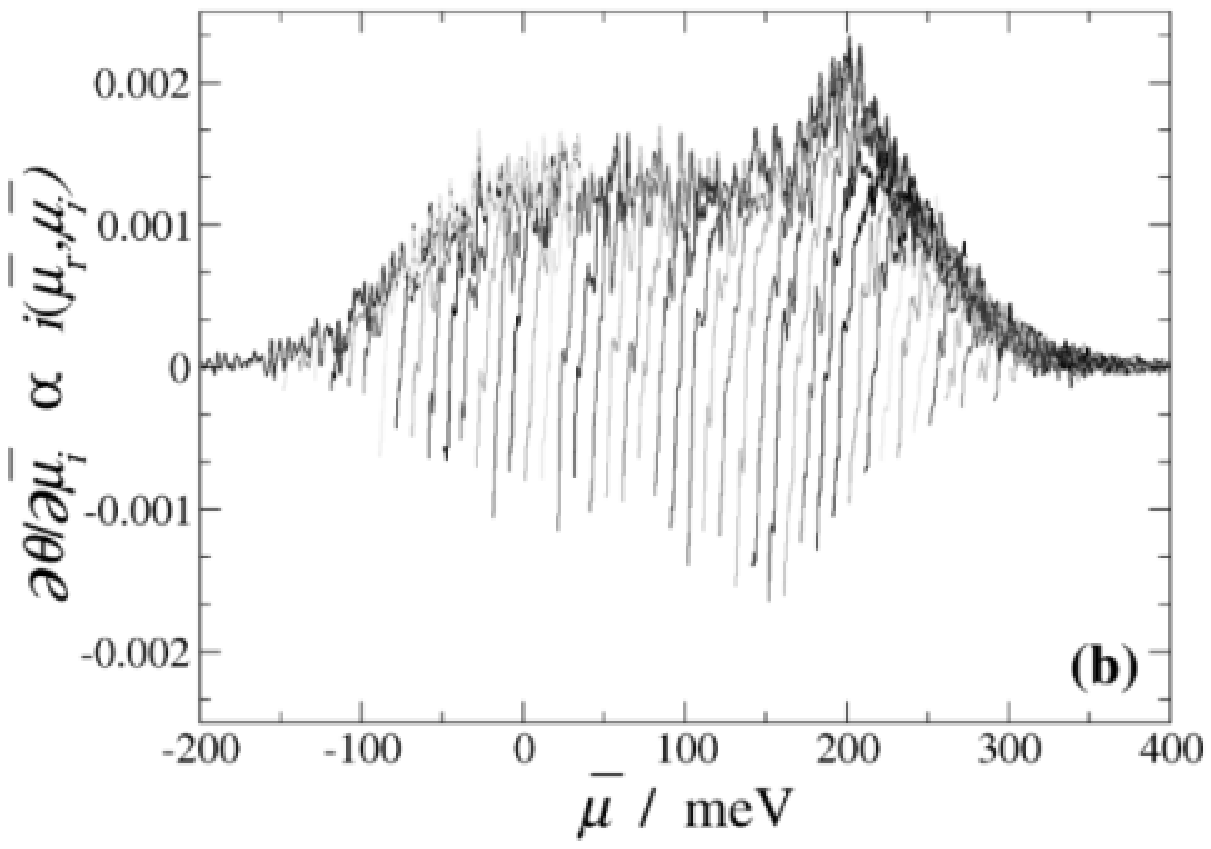}
	    \mbox{}
	    \vspace{0.1in}
	    \mbox{}
	    \includegraphics[width=.48\textwidth]{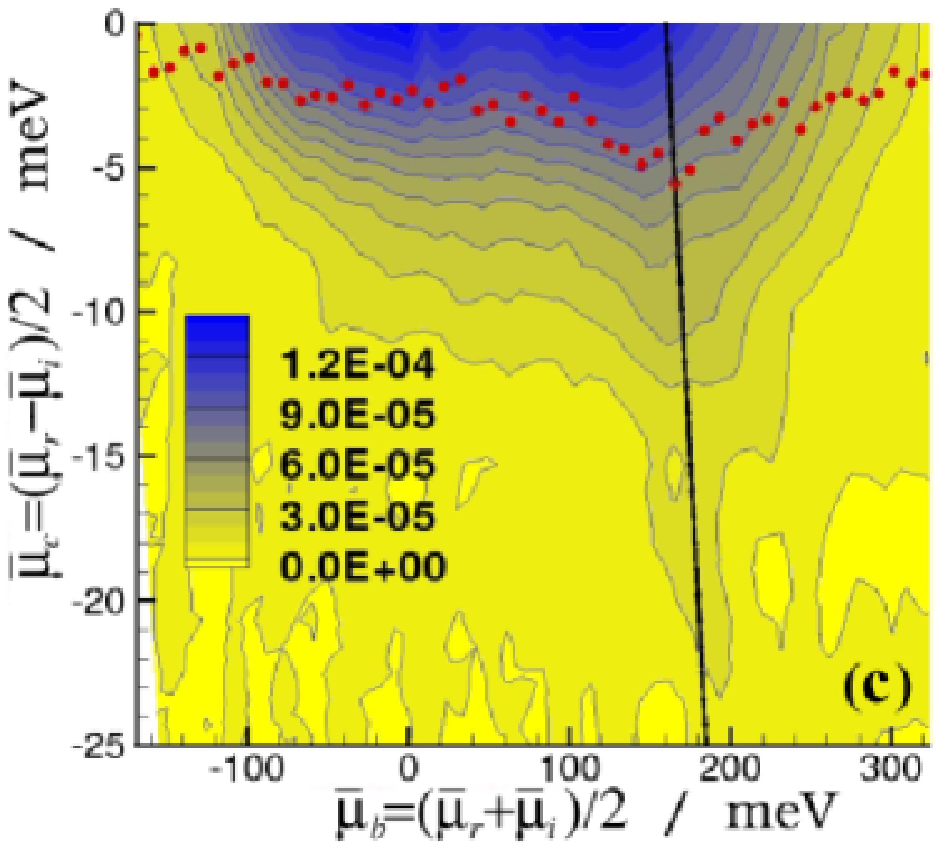}
	  \end{center}
	  \caption[]{ (Color online.)
	    ({\bf a})
	    FORCs for a continuous phase
	    transition simulated at a slow scan rate, $\Omega=0.0003$\,meV/MCSS.
	    The thin black curve near the middle of the major loop
	    is the equilibrium isotherm. The inset
	    is a magnification of the critical region. The minimum 
	    of each FORC is also shown (filled circles). The thick,
     	    black curve shows the first FORC which dips below the critical
	    coverage, $\theta_c \approx 0.36$.
({\bf b})
The corresponding voltammetric currents, calculated by
numerical differentiation of the FORCs. See Eq.~\ref{eq:curr}. 
	    ({\bf c})
	    FORC diagram generated from the FORCs in ({a}). The
	    positions of the FORC minima are shown as filled circles. The
	    thick, straight line corresponds to the FORC marked as a
	    thick curve in ({a}). 
%After Ref.~\protect\cite{AbouHamad:06}.
	  }
	  \label{fig:fig2}
	\end{figure}
	Using the same Hamiltonian, but with long-range repulsive interactions
	and nearest-neighbor exclusion as appropriate for modeling
	halide electrosorption on Ag(100) 
	\cite{AbouHamad:04,MitchellSS:01,AbouHamad:03,AbouHamad:05},
	KMC simulations were
	used to produce the family of FORCs for a continuous phase
	transition. The reversal potentials $\bar{\mu}_r$ were
	separated by $\Delta \bar{\mu}_r = 10$\,meV 
	increments in the interval $[-200\,{\rm meV},400\,{\rm meV}]$. As in 
	Refs.~\cite{AbouHamad:04,AbouHamad:03,AbouHamad:05}, the repulsive
	$1/r^3$ interactions, with nearest-neighbor exclusion and
	$\phi_{\rm nnn} = -21$\,meV, are calculated with exact
	contributions for $r_{ij} \le 3$, and using a mean-field
	approximation for $ r_{ij} > 3$. The barriers for
	adsorption/desorption and nearest- and next-nearest-neighbor
	diffusion were approximated on the basis of DFT calculations 
	\cite{Mitchell:02} as $\Delta_{\rm a/d} = 300$\,meV, $\Delta_{\rm nn}
	= 100 $\,meV, and $\Delta_{\rm nnn} = 200$\,meV,
	respectively \cite{AbouHamad:04}. Larger values of the
	diffusion barrier were also used to study the effect of
	diffusion on the dynamics. A continuous phase transition
	occurs between a disordered state at low coverage and an
	ordered state at high coverage \cite{Ocko:97,Wandlowski:01}.
        The FORCs, voltammetric currents, and FORC diagram are shown in
	Fig.~\ref{fig:fig2}. 
	
	Also indicated in
	Fig.~\ref{fig:fig2}({a}) are the FORC minima and the
	equilibrium isotherm, as calculated in independent equilibrium
	Monte Carlo simulations.
	Note that the FORC minima in Fig.~\ref{fig:fig2}({a}) 
        lie directly on the equilibrium
	isotherm. This is because such a system has one stable state
	for any given value of the potential, as defined by the
	continuous, single-valued 
	equilibrium isotherm. 
The corresponding voltammetric currents are shown in
Fig.~\ref{fig:fig2}(b). 
        The uniformly positive value of the FORC
	diagram in Fig.~\ref{fig:fig2}({c}) reflects the convergence
	of the family of FORCs with increasing $\bar{\mu}_i$. This convergence
	results from relaxation toward the equilibrium isotherm, at a
	rate which increases with the distance from equilibrium. It is
	interesting to note that, while it is difficult to see at this
	slow scan rate, the rate of approach to equilibrium
	decreases greatly along the first FORC that dips below the
	critical coverage $\theta_c \approx 0.36$ (shown in bold in
	Fig.~\ref{fig:fig2}({a})). The FORCs that lie completely
	in the range $\theta > \theta_c$ never enter into the disordered
	phase, and thus their approach to equilibrium is not hindered
	by jamming. This is a phenomenon that occurs when further
	adsorption in a disordered adlayer is hindered by the
	nearest-neighbor exclusion. As a result, extra diffusion steps
	are needed to make room for the new adsorbates, and the system
	follows different dynamics than a system with an ordered
	adlayer \cite{Privman:93}. The FORCs that dip below
	$\theta_c$ enter into the disordered phase, and thus
	their approach to equilibrium is delayed by jamming. This is
	reflected in the FORC diagram by the Florida-shaped
	``peninsula'' centered around this FORC in
	Fig.~\ref{fig:fig2}({c}).

\begin{figure}
	  \vspace{0.4truecm}
	  \begin{center}
	    \includegraphics[width=.5\textwidth]{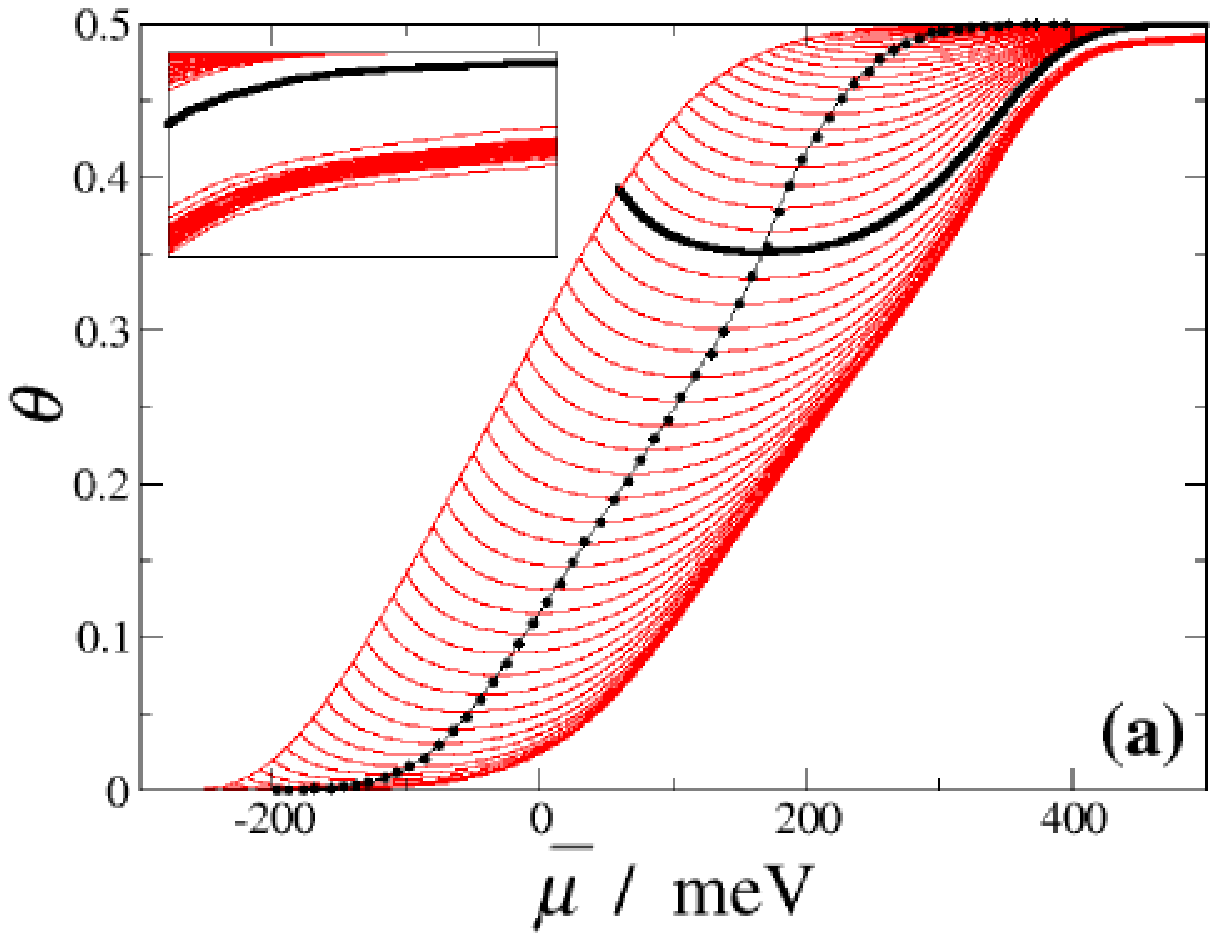}
	    \mbox{}
	    \vspace{0.2in}
	    \mbox{}
	    \includegraphics[width=.5\textwidth]{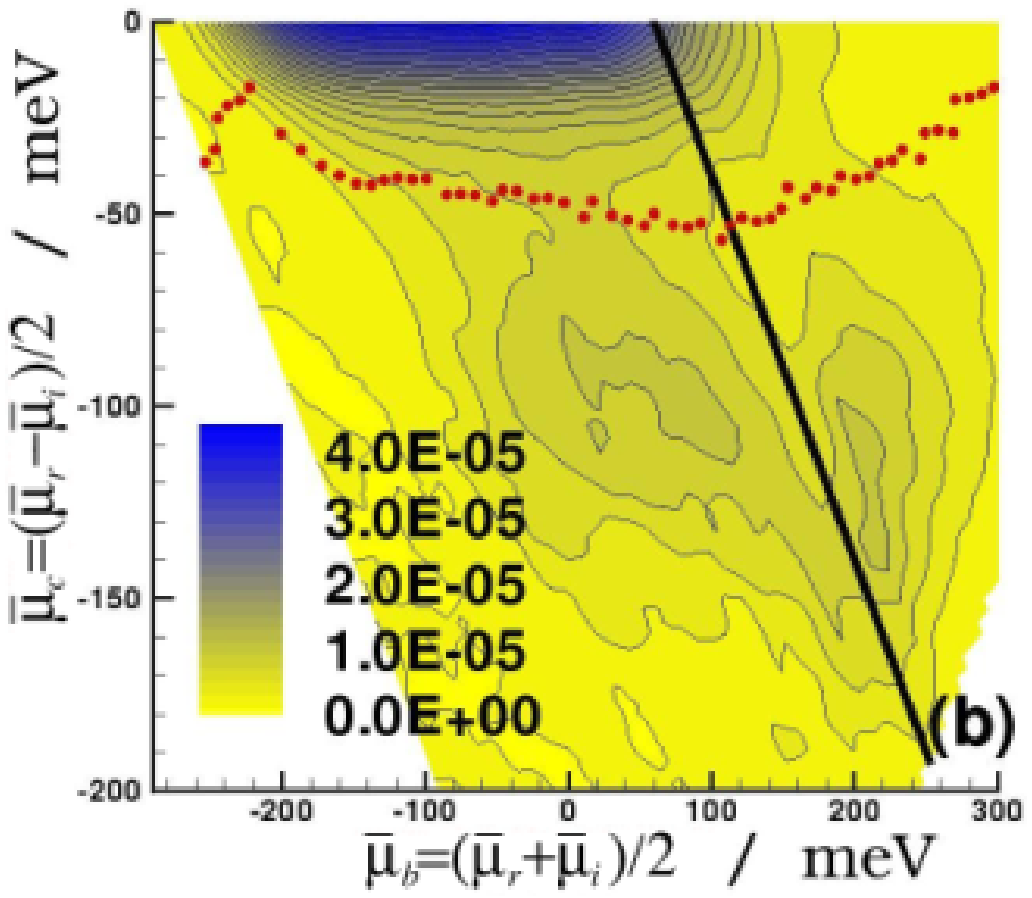}
	  \end{center}
	  \caption[]{ (Color online.)
	    ({\bf a})
	    FORCs for a continuous phase
	    transition simulated at a high scan rate,
	    $\Omega=0.01$~meV/MCSS, and the other model parameters as
	    given in the text. 
	    The different curves and symbols have the same meanings
	    as in Fig.~\protect\ref{fig:fig2}(a).
	    The inset showing the region of large $\bar{\mu}$ and
	    $\theta$ emphasizes the jamming behavior. 
	    ({\bf b})
	    FORC diagram generated from the FORCs in ({\bf a}). 
	    The different lines and symbols have the same meanings
	    as in Fig.~\protect\ref{fig:fig2}(b). 
	  }
	  \label{fig:highscan}
	\end{figure}

\begin{figure}
	  \vspace{0.4truecm}
	  \begin{center}
	    \includegraphics[width=.5\textwidth]{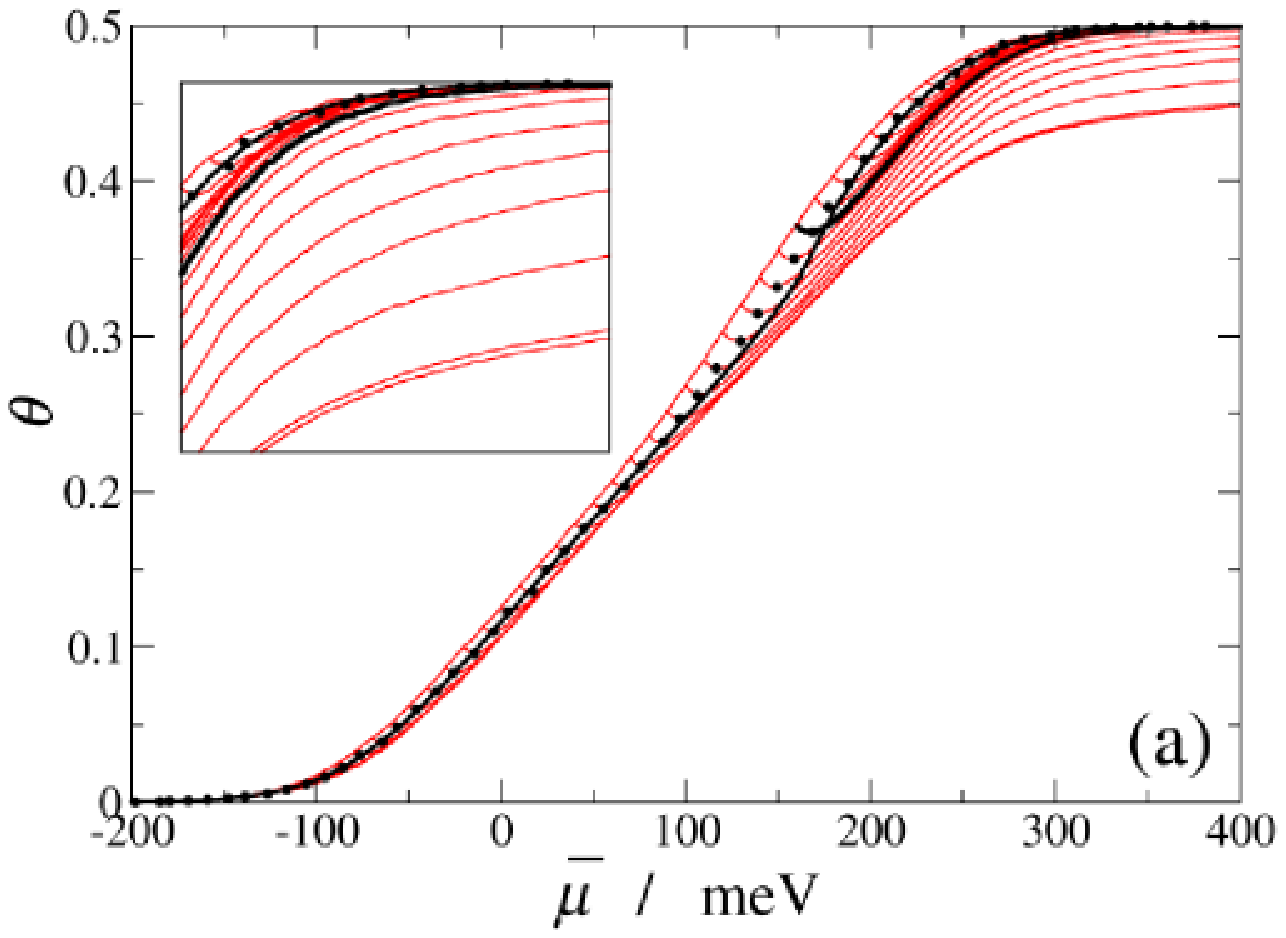}
	    \mbox{}
	    \vspace{0.2in}
	    \mbox{}
	    \includegraphics[width=.5\textwidth]{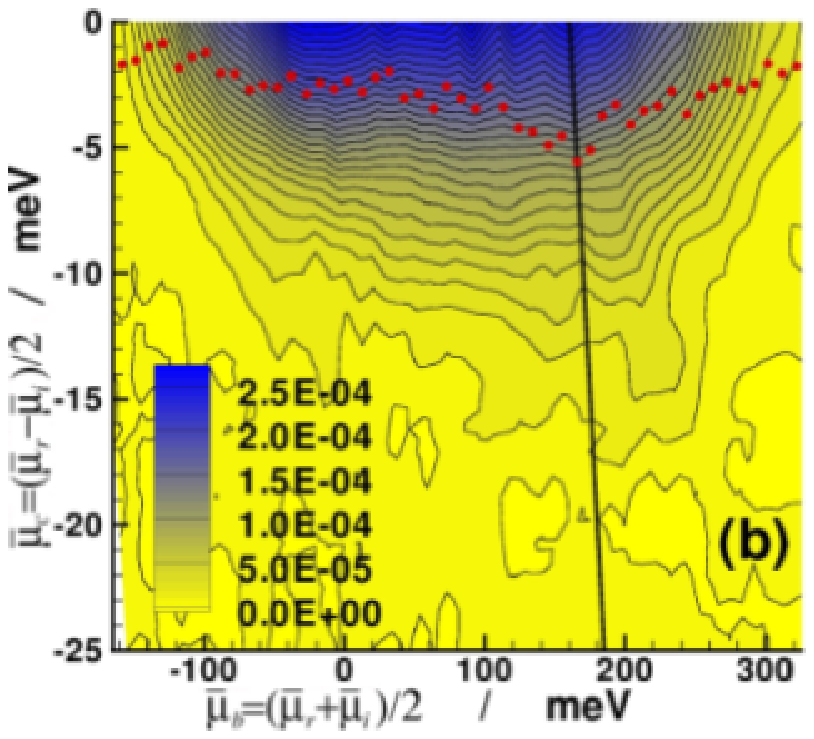}
	  \end{center}
	  \caption[]{ (Color online.)
	    ({\bf a})
	    FORCs for a continuous phase
	    transition simulated with $\Omega = 0.0003$~meV/MCSS and 
	    a large diffusion barrier, $\Delta_{\rm nn}=300$~meV.
	    The other model parameters are as given in the text.
	    The different curves and symbols have the same meanings
	    as in Figs.~\protect\ref{fig:fig2}(a) and
	    \protect\ref{fig:highscan}(a).
	    The inset showing the region of large $\bar{\mu}$ and
	    $\theta$ emphasizes the jamming behavior. 
	    ({\bf b})
	    FORC diagram generated from the FORCs shown in ({\bf a}). 
	    The different lines and symbols have the same meanings
	    as in Figs.~\protect\ref{fig:fig2}(b) and 
	    \protect\ref{fig:highscan}(b). 
	  }
	  \label{fig:lowdiff}
	\end{figure}

        The effect of jamming is more
	pronounced at higher scan rates, or with a higher diffusion
	barrier, where the rate of adsorption is much faster than the
	rate of diffusion. The family of FORCs and FORC diagram at a
	higher scan rate, $\Omega=0.01$\,meV/MCSS, are shown in
	Fig.~\ref{fig:highscan}, and with a larger diffusion barrier in
	Fig.~\ref{fig:lowdiff}. In Fig.~\ref{fig:highscan}, two
	distinct groups of FORCs undergoing jammed and unjammed
	dynamics can be clearly seen. This is reflected in the FORC
	diagram as a splitting of the ``peninsula'' into two
	``islands'' of locally maximal values of $\rho$. A similar effect is
	seen in Fig.~\ref{fig:lowdiff}, since also there the rate of
	adsorption is much faster than the rate of diffusion (larger
	diffusion barrier). In addition, Fig.~\ref{fig:lowdiff}({a}) shows a
	slight difference between the FORC minima and the equilibrium
	curve around the critical coverage. Notice also in
	Fig.~\ref{fig:highscan}({a}) that even at a much higher
	scan rate than in Fig.~\ref{fig:fig2} (nearly two orders of
	magnitude), the FORC minima still follow the equilibrium curve
	very accurately. Thus, the EC-FORC method should be useful to
	obtain the equilibrium adsorption isotherm quite accurately
        in experimental systems with slow equilibration rates.
	
	\section{Comparison and conclusions}
	\label{sec:conc}
	Two observations can be made by comparing the FORCs, 
	voltammetric currents, and FORC
	diagrams for systems with discontinuous and continuous phase
	transitions. First, the FORC minima in systems with a
	continuous phase transition correspond closely to the equilibrium
	behavior, while they do not for systems with a discontinuous
	phase transition. Thus, FORCs can be used to recover the
	equilibrium behavior for systems with continuous phase
	transitions that need a long time to equilibrate. This should
	be useful in experiments. Second, due to the instability that
	exists in systems with a discontinuous phase transition, the
	minima of the family of FORCs in this case 
	form a back-bending ``van der
	Waals loop,'' and the corresponding FORC diagram contains
	negative regions that do not exist for systems with a
	continuous phase transition. Since experimental implementation
	of the EC-FORC method should only require simple reprogramming of
	a potentiostat designed to carry out a standard CV experiment, we
	believe the method can be of significant use in obtaining
	additional dynamic as well as equilibrium information from
	such experiments for systems that exhibit electrochemical
	adsorption with related phase transitions.
	
	\section*{Acknowledgments}
We gratefully acknowledge useful comments from E.~Borguet and two
anonymous referees. 
	
	This research was supported by U.S. NSF Grant
	No. DMR-0240078, and by Florida State University through the School of 
	Computational Science, the Center for Materials Research and
	Technology, and the National High Magnetic Field Laboratory.

\end{document}